\documentclass[conference]{IEEEtran}
\IEEEoverridecommandlockouts

\usepackage{cite}
\usepackage{amsmath,amssymb,amsfonts}
\usepackage{graphicx}
\usepackage{textcomp}
\usepackage{xcolor}
\usepackage{booktabs}
\usepackage{hyperref}  
\usepackage[ruled,vlined]{algorithm2e} 
\usepackage{xcolor} 
\usepackage{graphicx} 
\usepackage{balance} 

\graphicspath{ {figures/} }
\def\BibTeX{{\rm B\kern-.05em{\sc i\kern-.025em b}\kern-.08em
    T\kern-.1667em\lower.7ex\hbox{E}\kern-.125emX}}
\begin{document}

\title{SmartAttack: Air-Gap Attack via Smartwatches}

\author{\IEEEauthorblockN{Mordechai Guri}
\IEEEauthorblockA{\textit{Ben-Gurion University of the Negev} \\
\textit{Beer Sheva, Israel} \\
gurim@post.bgu.ac.il \\ \url{https://www.covertchannels.com/}}
}

\maketitle

\begin{abstract}	
Air-gapped systems are considered highly secure against data leaks due to their physical isolation from external networks. Despite this protection, ultrasonic communication has been demonstrated as an effective method for exfiltrating data from such systems. While smartphones have been extensively studied in the context of ultrasonic covert channels, smartwatches remain an underexplored yet effective attack vector.

In this paper, we propose and evaluate \textit{SmartAttack}, a novel method that leverages smartwatches as receivers for ultrasonic covert communication in air-gapped environments. Our approach utilizes the built-in microphones of smartwatches to capture covert signals in real time within the ultrasonic frequency range of 18–22~kHz. Through experimental validation, we assess the feasibility of this attack under varying environmental conditions, distances, orientations, and noise levels. Furthermore, we analyze smartwatch-specific factors that influence ultrasonic covert channels, including their continuous presence on the user's wrist, the impact of the human body on signal propagation, and the directional constraints of built-in microphones. Our findings highlight the security risks posed by smartwatches in high-security environments and outline mitigation strategies to counteract this emerging threat.
\end{abstract}


\section{Introduction}
Air-gapped systems are commonly deployed in high-security environments to protect sensitive information. Beyond air-gapped systems, many highly secured networks rely on strict monitoring and Data Leakage Prevention (DLP) solutions to prevent unauthorized data transfers. These systems employ network traffic analysis, endpoint security controls, and strict access policies to detect and block suspicious exfiltration attempts \cite{zander2007survey, guri2018bridgeware}. In such environments, attackers cannot rely on conventional exfiltration methods (e.g., Internet access) and must instead exploit non-standard covert channels that bypass traditional networking mechanisms. Techniques such as acoustic signals, power fluctuations, and peripheral device manipulation provide alternative pathways for stealthy data extraction from secured systems \cite{carrara2016out,na2024study}.

\subsection{Air-Gap Covert Channels}
Previous research has identified various covert communication channels that can be exploited for data exfiltration, including electromagnetic, optical, thermal, and acoustic channels \cite{carrara2016out,guri2018bridgeware}. Among these, ultrasonic channels have received significant attention due to their stealthiness and the widespread presence of microphones and speakers in modern devices \cite{carrara2015acoustic}.

Ultrasonic communication typically involves a transmitter and a receiver. The transmitter, such as a PC or laptop with a speaker, generates ultrasonic sound waves that remain inaudible to humans, typically at frequencies of 18 kHz and above \cite{carrara2015acoustic}. The receiver, a nearby device equipped with audio recording capabilities, detects and captures these signals. An attacker can encode sensitive data, including keystrokes, encryption keys, credentials, biometric information, and confidential documents, into ultrasonic waves and covertly transmit them to the receiver without the user's awareness.

\subsection{Smartwatches as Covert Communication Receivers}
Smartphones and tablets have been widely studied as both transmitters and receivers in ultrasonic covert channels. However, smartwatches—despite their growing popularity and frequent presence in high-security environments—remain largely unexplored in this context \cite{khan2024vulnerability}.

The portability and wearable nature of smartwatches significantly enhance their potential as covert communication receivers. These devices are designed for continuous wear, allowing them to be discreetly carried into various environments, including secured or air-gapped systems. Their seamless integration into daily life ensures proximity to potential signal sources, thereby reducing the transmission distance required for ultrasonic communication. Additionally, the unobtrusive design of smartwatches allows them to operate without drawing attention, a crucial factor in covert cyberattacks \cite{lee2016security}.

\subsection{Smartwatches Characteristics}
Unlike smartphones, smartwatches possess unique characteristics that influence their role in ultrasonic covert channels and require careful evaluation. Unlike stationary smartphones, smartwatches are worn on the wrist, leading to continuous movement that affects signal stability. The orientation of the wrist relative to the transmitting computer impacts signal reception, while the presence of the human body introduces attenuation and signal distortion. These factors, combined with the smartwatch's smaller microphone and lower processing power compared to smartphones, create distinct challenges that necessitate specialized analysis.

The contribution of this paper is as follows: We explore the feasibility of using smartwatches for ultrasonic covert communication attacks, detailing the attack model, implementation, and evaluation. We systematically assess their effectiveness as receivers by analyzing performance under varying distances, orientations, and transmitter types.

\section{Related Work}
Air-gapped systems have long been considered highly secure against remote attacks because they are physically isolated from unsecured networks. However, various covert channels have been identified that can bridge this isolation, enabling unauthorized data exfiltration. These channels exploit different physical phenomena, including electromagnetic, optical, magnetic, vibration-based, and thermal methods \cite{carrara2016out,guri2018bridgeware}. 

Electromagnetic covert channels manipulate electromagnetic emissions from computer components to transmit data. For instance, the AirHopper attack demonstrates data exfiltration from an air-gapped computer to a nearby mobile phone using FM radio signals \cite{guri2014airhopper}. Optical covert channels utilize light-emitting components for communication. The LED-it-GO technique controls a computer’s hard drive indicator LED to transmit data to an external observer \cite{guri2017led}. Magnetic covert channels exploit the magnetic fields generated by a computer's CPU, allowing data transmission to nearby devices equipped with magnetic sensors. Techniques such as ODINI facilitate this type of exfiltration \cite{guri2019odini}. Vibration-based covert channels leverage malware to induce vibrations in a device’s components, which can be detected by accelerometers in nearby devices, enabling data transfer via physical vibrations \cite{guri2021exfiltrating}. Thermal covert channels, such as BitWhisper, induce temperature changes in a computer, which are then detected by thermal sensors in adjacent systems, facilitating bidirectional communication \cite{guri2015bitwhisper}.

\subsection{Acoustic Covert Channels}
Acoustic communication relies on sound waves to transmit information. While audible sound ranges from 20 Hz to 20 kHz, ultrasonic frequencies exceed 20 kHz, making them inaudible to humans. Ultrasonic communication utilizes these high-frequency sound waves, which can be generated and received by speakers and microphones capable of handling such signals. 

The use of sound as a communication medium dates back to early technologies like line modems and fax machines, which relied on acoustic signals for data transmission \cite{hanes2008fax}. These systems encoded information into both audible and inaudible frequencies, laying the groundwork for modern sound-based communication. In recent years, the prevalence of microphones and speakers in everyday devices has enabled new applications, including seamless device pairing and covert data transmission \cite{hanspach2014recent}.  

Previous studies have demonstrated various ultrasonic covert communication techniques. Zhang et al. \cite{zhang2022ultrannel} introduced Ultrannel, an air-gap covert channel that leverages ultrasonic waves through capacitor-microphone interactions. Wong demonstrated the feasibility of ultrasonic communication between air-gapped systems using speakers and microphones, creating a covert command-and-control channel \cite{wong2018crossing}. Guri et al. investigated speaker-to-speaker ultrasonic communication, analyzing the effects of environmental noise on covert channel performance \cite{guri2020speaker}. Sherry et al. explored near-ultrasonic covert channels using software-defined radio (SDR) techniques, showing that malware can exploit ultrasonic frequencies for data exfiltration \cite{sherry2023near}. Carrara and Adams analyzed the effectiveness of acoustic covert channels in different environments, evaluating both ultrasonic and audible sound-based transmissions \cite{carrara2015acoustic}. Matyunin et al. proposed a vibrational covert channel, which utilizes low-frequency acoustic signals instead of ultrasonic waves \cite{matyunin2019vibrational}. Carrara and Adams also conducted a survey on out-of-band covert channels, categorizing various side-channel and device-pairing threats \cite{carrara2016out}. Guri et al. proposed Mosquito, a method for covert ultrasonic communication between air-gapped computers using speaker-to-speaker transmission \cite{guri2018mosquito}. 

Beyond covert channels, ultrasonic signals can also be exploited for direct command injection attacks, posing security risks to voice-controlled devices. DolphinAttack manipulates voice assistants like Siri and Alexa by embedding inaudible ultrasonic commands, enabling unauthorized actions without user awareness \cite{zhang2017dolphinattack}. Similarly, SurfingAttack transmits malicious commands via ultrasonic waves through solid surfaces, allowing remote control of nearby smartphones without direct contact \cite{yan2020surfingattack}. These attacks highlight the broader security implications of ultrasonic signal exploitation beyond covert communication.

\subsection{Smartwatches}
Table~\ref{tab:acoustic_channels} summarizes various studies on acoustic communication channels, outlining the types of transmitters, receivers, and frequency bands used in covert communication research. While smartphones have been extensively studied in the context of ultrasonic covert communication, smartwatches remain largely unexplored \cite{khan2024vulnerability}. Given their widespread adoption and constant proximity to users, smartwatches present a unique opportunity for covert data exfiltration. Our research focuses on utilizing the built-in microphones and speakers of smartwatches to establish ultrasonic covert channels, assessing their feasibility and security implications.

\begin{table}[]
	\centering
	\caption{Summary of different works on acoustic communication channels, detailing the transmitter, receiver, and frequency bands used.}
	\label{tab:acoustic_channels}
	\begin{tabular}{@{}llll@{}}
		\toprule
		Ref. & Transmitter        & Receiver   & Bands                    \\ \midrule
		\cite{carrara2015acoustic,guri2018mosquito}    & Laptop/PC             & Laptop/PC    & Ultrasonic               \\
		\cite{guri2023grillo,wong2018crossing} & Workstation        & Smartphone & Ultrasonic / Sonic       \\
		\cite{matyunin2019vibrational,guri2021exfiltrating}    & Workstation        & Smartphone & Infrasonic / Vibrational \\
		\cite{deshotels2014inaudible, pandya2022shoutimei}    & Smartphone         & Smartphone & Ultrasonic               \\
		SmartAttack    & Workstation/Laptop & Smartwatch & Ultrasonic / Sonic       \\ \bottomrule
	\end{tabular}
\end{table}

\section{Attack Model with Smartwatches}
Smartwatches, by virtue of their design and usage, are frequently in close proximity to computers, making them particularly effective as receivers for ultrasonic signals. Their inherent portability ensures that they accompany users into various environments, including secured spaces where computers are present. This constant movement between different locations enhances their potential utility in covert communication scenarios. Additionally, the seamless integration of smartwatches into daily life allows them to passively gather signals without drawing attention, a critical factor for discreet data transmission. 

The proposed attack model exploits smartwatches as covert ultrasonic receivers for data exfiltration from secured, potentially air-gapped networks. The attack consists of several key stages, described below.

\subsection{Infiltration}
To initiate the attack, the adversary must first infiltrate the secured network, which may be air-gapped to prevent unauthorized data leakage. Despite its isolation, past cybersecurity incidents have shown that air-gapped networks are not impervious to compromise \cite{park2023survey,na2024study}. Attackers have successfully breached such systems through methods such as supply chain attacks, insider threats, or the use of infected removable media like USB drives. Once the malware is implanted within the secure network, it remains dormant or operates stealthily, awaiting the next stages of the attack \cite{sati2023analysis,lindsay2013stuxnet}.

The next step involves compromising a smartwatch belonging to a visitor or an employee with access to the secured environment. Modern smartwatches run advanced operating systems and are equipped with multiple connectivity options, including Wi-Fi, Bluetooth, NFC, and email applications \cite{king2018survey}. These capabilities make them susceptible to malware infections via malicious applications, phishing attacks, or exploits targeting wireless communication protocols. Once compromised, the smartwatch malware operates covertly, continuously monitoring its environment for incoming ultrasonic signals.

The malware on the compromised computer is responsible for gathering sensitive information such as keystrokes (keylogging), encryption keys, biometric data, or user credentials. This information is then modulated onto ultrasonic signals in the inaudible frequency range (18 kHz and above). Using the computer’s speakers, the malware transmits these covert signals, leveraging ultrasonic propagation to evade human detection.

\subsection{Exfiltration}
Simultaneously, the infected smartwatch continuously scans the acoustic spectrum for covert ultrasonic signals. Upon detecting a transmission, it demodulates and decodes the exfiltrated data, reconstructing the stolen information. The smartwatch then forwards the extracted data to the attacker using available communication channels such as Wi-Fi, cellular networks, or Bluetooth tethering, effectively bypassing traditional security measures. 

\subsection{Smartwatches as Ultrasonic Receivers}
Smartwatches possess several technological features that enable them to receive ultrasonic signals effectively. One key component facilitating this capability is the presence of high-sensitivity microphones capable of capturing frequencies beyond the human hearing range \cite{gras2024microphone}. Additionally, the integration of advanced signal processing software allows these devices to decode ultrasonic transmissions efficiently, enhancing their utility in covert communication scenarios \cite{soundwatch2024}. The processing units within smartwatches further support real-time analysis of ultrasonic signals, making them well-suited for discreet data reception. Moreover, their compact and ergonomic design ensures they remain unobtrusive while performing these functions, making them an ideal tool for receiving ultrasonic signals in varied environments.

\begin{figure}[]
	\centering
	\includegraphics[width=1\linewidth]{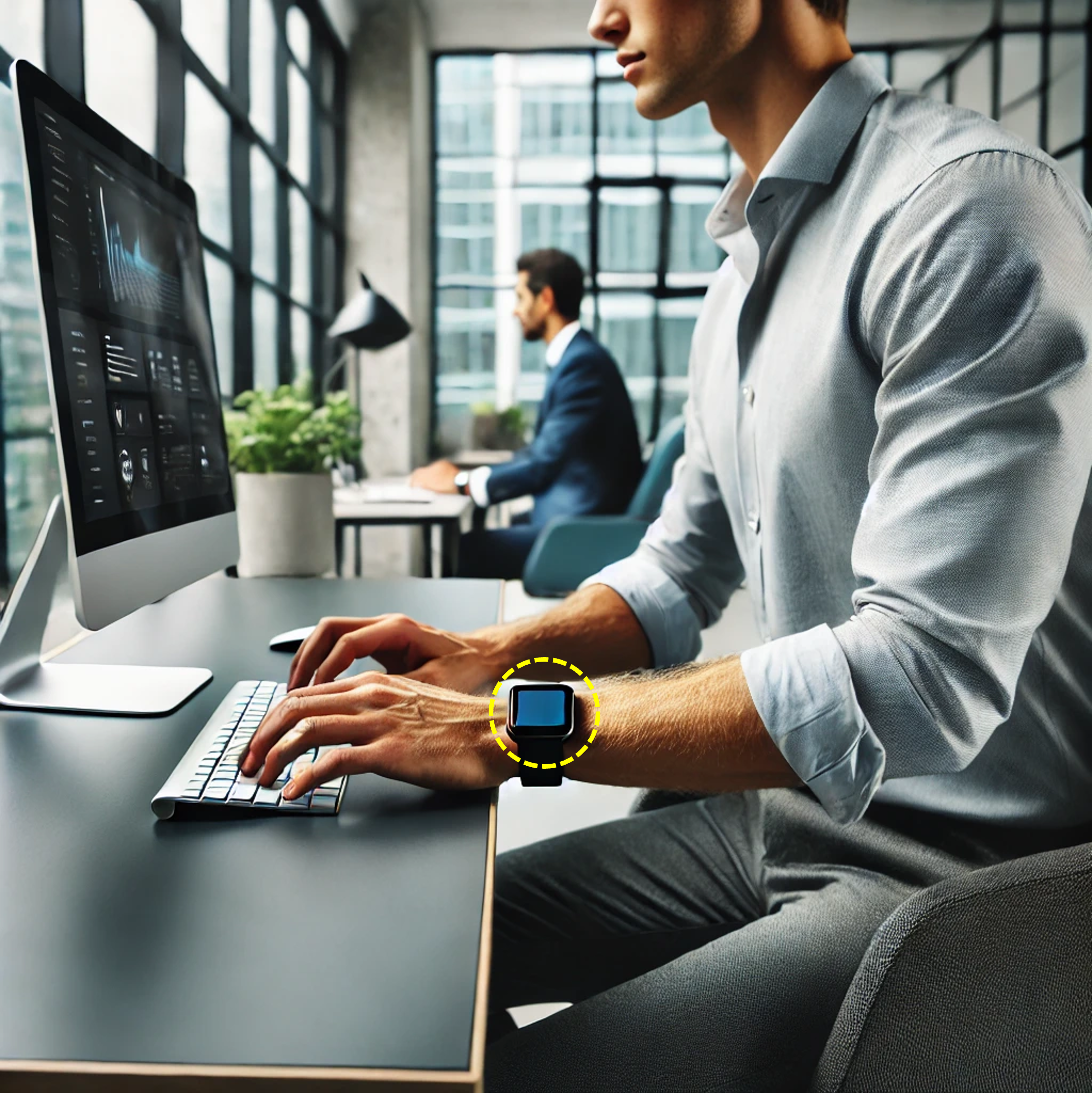}
	\caption{Illustration of the attack model. A programmer working in front of a highly secure or air-gapped computer wears a smartwatch on their wrist. The infected computer transmits sensitive data, such as keylogging information, modulated onto ultrasonic signals, which are covertly received by the smartwatch.}
	\label{fig:attack_model}
\end{figure}

Figure~\ref{fig:attack_model} illustrates the attack scenario, where a programmer is seated in front of a highly secure or air-gapped computer, wearing a smartwatch on their wrist. The compromised computer transmits sensitive information, such as keystrokes, modulated onto ultrasonic signals, which are received and processed by the smartwatch.

\section{Design \& Implementation}
To evaluate the proposed method, we implemented a PC-based transmitter and a smartwatch receiver application. The acoustic covert communication system consists of two primary components: a transmitter, responsible for generating and modulating ultrasonic signals, and a Wear OS smartwatch receiver, which captures, demodulates, and decodes the transmitted data.

\begin{table}[b]
	\centering
	\caption{Comparison of Smartphones and Smartwatches for Ultrasonic Data Exfiltration}
	\label{tab:comparison}
	\resizebox{\columnwidth}{!}{ 
		\begin{tabular}{l l l} 
			\toprule
			\textbf{Factor} & \textbf{Smartphone} & \textbf{Smartwatch} \\
			\midrule
			Positioning & Desk, pocket, or hand & Worn on wrist, constantly moving \\
			Signal Reception & More stable placement & Affected by wrist movement \\
			Audio Hardware & High-quality mic \& DSP & Weaker mic, limited DSP \\
			Processing Power & Strong CPU/GPU & Lower processing power \\
			Signal Attenuation & Minimal obstructions & Affected by body absorption \\
			Stealth & Requires placement & Always worn, less noticeable \\
			Energy Constraints & Larger battery & Smaller battery, power-limited \\
			\bottomrule
		\end{tabular}
	} 
\end{table}

\subsection{Smartwatches vs. Smartphones Considerations}
Ultrasonic covert channels have been widely studied in the context of smartphones, but smartwatches introduce unique challenges and advantages in such attacks. Unlike smartphones, which are typically placed on a desk, in a pocket, or held in hand, smartwatches are worn on the wrist, resulting in continuous movement. This movement influences reception quality due to variations in distance \(d\), orientation \(\theta\), and body-induced signal attenuation \(\alpha\). As the wrist moves, the received signal strength \( S_r \) fluctuates according to the path loss model:

\[
S_r \propto d^{-\beta}
\]

where \( \beta \) is the path loss exponent, which depends on environmental conditions.

Another crucial factor is the hardware disparity between smartphones and smartwatches. Smartwatches generally feature microphones with a lower signal-to-noise ratio (SNR) and reduced digital signal processing (DSP) capabilities, making ultrasonic demodulation more difficult. Moreover, human body absorption introduces additional attenuation \(\alpha\), reducing received signal power. Despite these limitations, smartwatches offer significant advantages in covert communication due to their constant presence on the user's wrist, minimizing detection risks. A comparative analysis of smartwatches and smartphones as attack vectors is presented in Table~\ref{tab:comparison}.

\subsection{PC-Based Transmitter Implementation}
The transmitter is implemented in Java, utilizing real-time audio synthesis to generate ultrasonic signals within the inaudible frequency range. It employs Binary Frequency Shift Keying (B-FSK) modulation, allowing the selection of frequency pairs for transmission. For testing, we use a lower frequency $ f_0 = 18.5 $ kHz to represent binary ``0'' and a higher frequency $ f_1 = 19.5 $ kHz to represent binary ``1.'' Each bit is transmitted within a certain symbol duration, e.g., $T_s = 50$ ms, to ensure smooth frequency transitions and minimize spectral artifacts. To structure the transmission, each message begins with a preamble sequence, an alternating bit pattern of "101010," which aids in synchronization. The 16-bit payload data follows, modulated using B-FSK to ensure each bit is represented by its corresponding frequency. To enhance reliability, an error-detection mechanism (checksum) is appended to the message.

The modulation process dynamically generates sinusoidal waveforms at the selected frequencies using Java’s \texttt{javax.sound.sampled} API. The waveform is streamed in real time to the PC’s speaker, effectively creating an ultrasonic acoustic channel for covert transmission. The power of the emitted signal \( P_t \) is optimized to remain within the speaker’s limitations while ensuring reliable reception at distances up to \( d_{\max} \), given by:

\[
d_{\max} \approx \left( \frac{P_t}{P_{\text{min}}} \right)^{\frac{1}{\beta}}
\]

where \( P_{\text{min}} \) is the receiver's minimum detectable power.

\subsection{Wear OS Smartwatch Receiver}

The smartwatch serves as the ultrasonic receiver, continuously monitoring the acoustic spectrum for incoming signals. The recorded audio is first filtered to isolate the ultrasonic band, typically between \( 18 \) kHz and \( 22 \) kHz, using a bandpass filter that eliminates lower-frequency environmental noise. A Fast Fourier Transform (FFT) with a sliding window is then applied to detect the dominant frequency components, classifying each frame as either \( f_0 \) or \( f_1 \) based on the peak frequency detected.
Following frequency classification, the extracted symbols are aligned to match the expected bit duration, ensuring proper demodulation. The system then searches for the predefined preamble sequence to achieve frame synchronization. Upon detecting a valid preamble, the smartwatch extracts the payload data and applies an integrity check, such as a cyclic redundancy check (CRC). Successfully decoded messages are then displayed on the smartwatch interface or transmitted to a paired device via Bluetooth or Wi-Fi.

\subsection{Decoder and Noise Mitigation Implementation}
To efficiently decode ultrasonic transmissions, the smartwatch employs an optimized signal processing pipeline focused on noise reduction and signal enhancement. It captures and buffers the microphone input, applying a Butterworth bandpass filter to suppress out-of-band noise while preserving the integrity of the transmission signal. 
Additionally, spectral subtraction is applied in the frequency domain to attenuate residual background noise components. The Fast Fourier Transform (FFT) is computed over overlapping Hamming-windowed segments, ensuring accurate frequency estimation while reducing spectral leakage. A peak detection algorithm is then used to classify the received frequencies as either \( f_0 \) or \( f_1 \), reconstructing the binary sequence while applying Kalman filtering to smooth out frequency fluctuations and compensate for Doppler shifts \(\Delta f\) induced by wrist movement.

\begin{algorithm}[]
	\small 
	\SetAlgoLined
	\SetKwInOut{Input}{Input}
	\SetKwInOut{Output}{Output}
	
	\caption{\textbf{Ultrasonic Decoder for Wear OS Smartwatch}}
	\label{alg:wearos_decoder}
	
	\Input{Microphone audio stream}
	\Output{Decoded binary data}
	
	\textcolor{blue}{\tcc{Initialize audio capture and sample rate}}
	Initialize microphone and set sampling rate\;
	Define frequency mapping: $ f_0 \rightarrow \text{Binary '0'}, f_1 \rightarrow \text{Binary '1'}$\;
	
	\textcolor{blue}{\tcc{Real-time decoding loop}}
	\While{Microphone capturing is active}{
		Capture audio in overlapping frames\;
		Compute FFT of each frame\;
		Identify dominant frequency peak\;
		
		\uIf{Peak frequency is near $ f_0 $}{
			Append \textbf{'0'} to buffer\;
		}
		\ElseIf{Peak frequency is near $ f_1 $}{
			Append \textbf{'1'} to buffer\;
		}
	}
	
	\textcolor{blue}{\tcc{Synchronization and data extraction}}
	Detect preamble pattern for synchronization\;
	\If{Valid preamble detected}{
		Extract data and verify integrity\;
		\If{No errors detected}{
			Convert binary sequence to a byte stream\;
			Output decoded message\;
		}
	}
\end{algorithm}

To efficiently decode ultrasonic transmissions, the smartwatch employs an optimized signal processing pipeline. Initially, it captures and buffers the microphone input, applying a bandpass filter to remove ambient noise. The FFT is computed over overlapping windows, allowing frequency components to be analyzed in real time. The receiver classifies detected frequencies as either \( f_0 \) or \( f_1 \), reconstructing the binary sequence while compensating for Doppler shifts \(\Delta f\) caused by wrist movement.

The decoding process is outlined in Algorithm~\ref{alg:wearos_decoder}, ensuring reliable symbol extraction and error resilience. Once the bitstream is reconstructed, the smartwatch identifies the synchronization preamble and extracts the valid payload data. If the integrity check confirms the absence of errors, the decoded information is converted into its original form.

\section{Evaluation \& Analysis}

We evaluate the impact of smartwatch orientation on ultrasonic signal reception to understand how positioning influences the reliability and effectiveness of covert communication channels. Specifically, we analyze the smartwatch’s ability to receive ultrasonic transmissions at different angular positions relative to the transmitting computer. The study considers eight orientations, denoted as \(\theta \in \{0^\circ, 45^\circ, 90^\circ, 135^\circ, 180^\circ, 225^\circ, 270^\circ, 315^\circ\}\). At each position, spectrograms are recorded to assess variations in signal strength, frequency response, and attenuation patterns.

\begin{figure}[]
	\centering
	\includegraphics[width=0.8\linewidth]{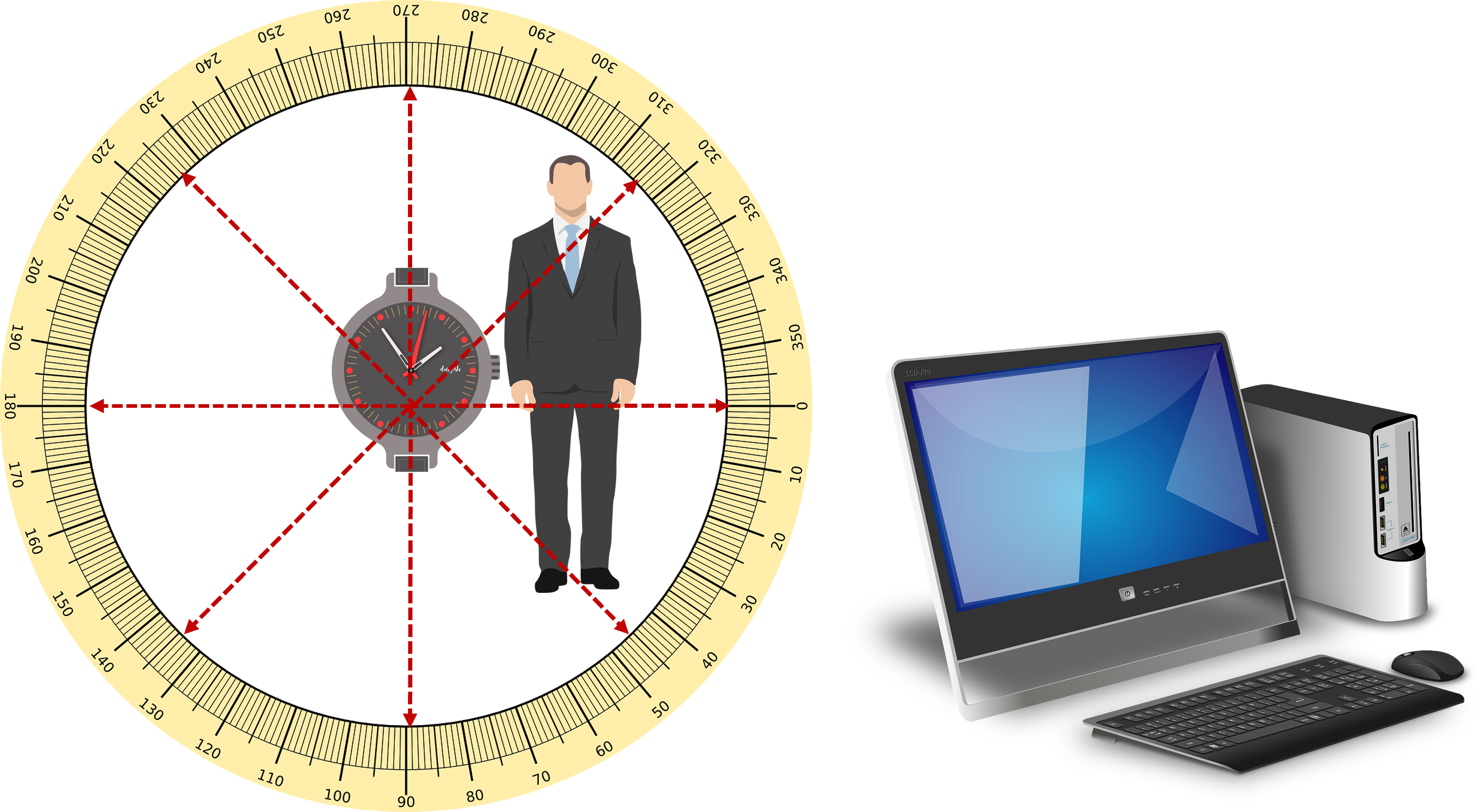}
	\caption{Illustration of smartwatch orientations at different angles relative to the transmitting computer.}
	\label{fig:illustration_angles}
\end{figure}

Figure~\ref{fig:illustration_angles} provides an illustration of the smartwatch orientations considered in our experiments. A spectrogram analysis is performed at each angle to investigate how signal reception is affected by factors such as directional microphone sensitivity, body occlusion, and environmental reflections. The results, shown in Figure~\ref{fig:spectrogram_angles}, demonstrate that orientation plays a crucial role in ultrasonic data transmission, significantly influencing reception quality.

\subsection{Directional Response Analysis of Smartwatch}

\begin{figure*}[h]
	\centering
	\includegraphics[width=0.9\textwidth]{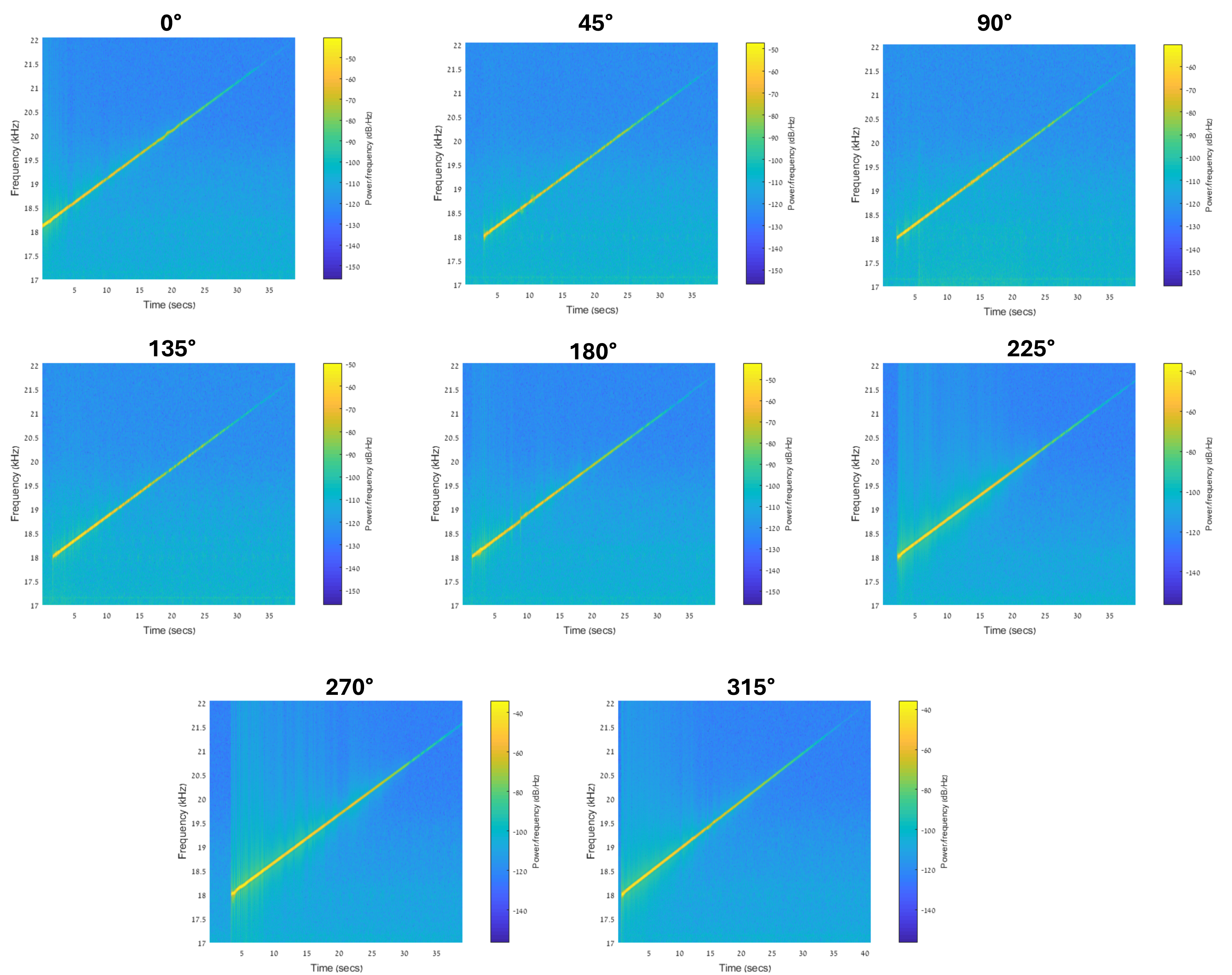}
	\caption{Spectrogram analysis of smartwatch response at different angles relative to the transmitting computer.}
	\label{fig:spectrogram_angles}
\end{figure*}

The spectrograms in Figure~\ref{fig:spectrogram_angles} demonstrate that signal reception is highly dependent on the smartwatch's orientation. When the smartwatch is positioned at \(\theta = 0^\circ\), the received signal is weakest, exhibiting maximum attenuation—\textit{in contrast to} the transmitter side. As \(\theta\) increases, the signal strength follows an attenuation function \( S_r(\theta) \), influenced by both body occlusion and the directional sensitivity of the smartwatch microphone. The strongest reception occurs in the range \( 180^\circ \leq \theta \leq 225^\circ \), where the smartwatch maintains a direct line-of-sight with the transmitting computer.

\begin{figure}[h]
	\centering
	\includegraphics[width=1\linewidth]{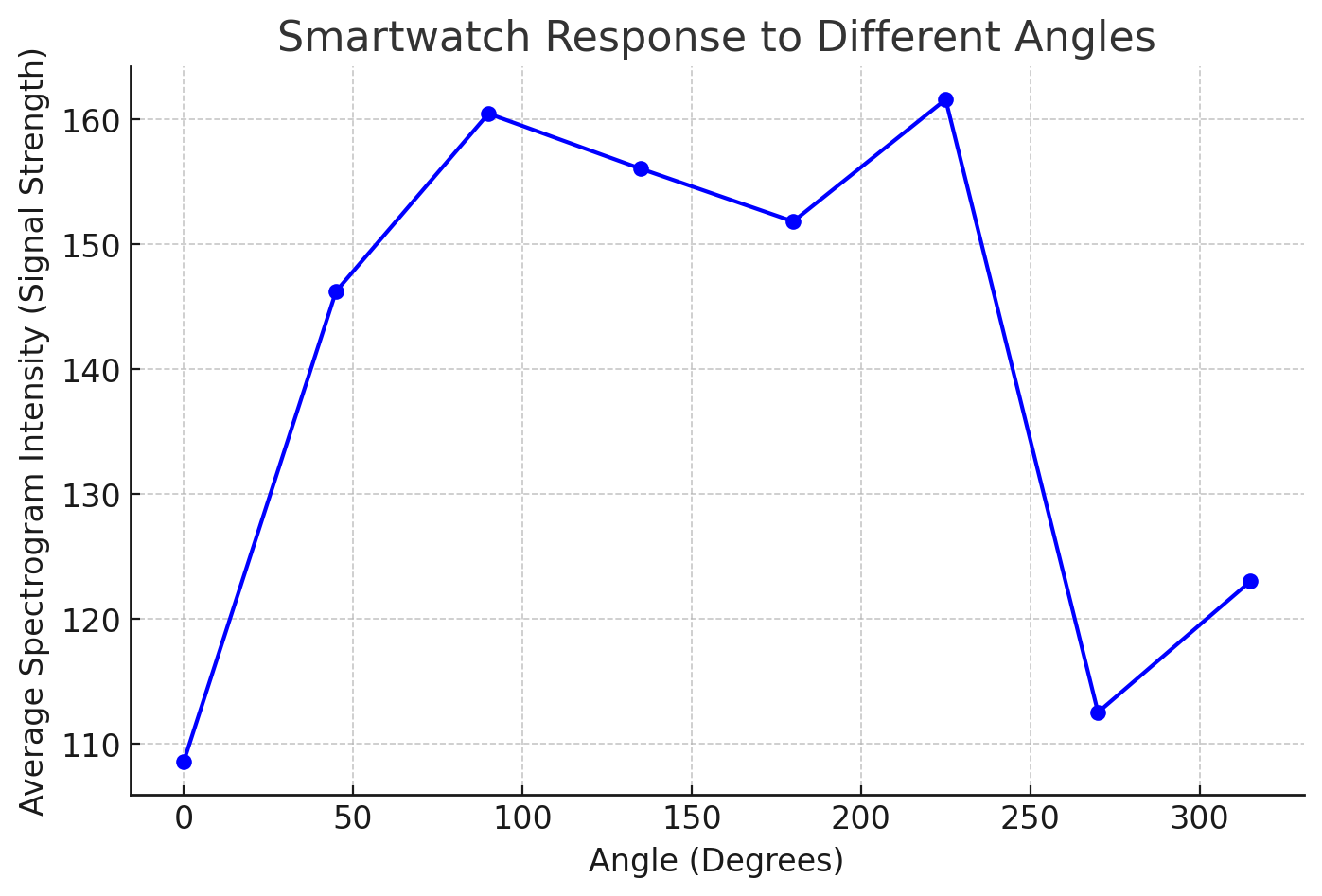}
	\caption{Frequency intensity profiles for different smartwatch orientations.}
	\label{fig:frequency_profiles}
\end{figure}

The frequency intensity profiles in Figure~\ref{fig:frequency_profiles} further illustrate how smartwatch orientation influences ultrasonic reception. The signal strength, denoted as \( I(f, \theta) \), varies as a function of both frequency \( f \) and orientation \( \theta \). At \(\theta = 90^\circ\) and \(\theta = 225^\circ\), the signal exhibits higher intensity across the ultrasonic spectrum, suggesting an optimal alignment between the smartwatch microphone and the transmitter. Conversely, orientations at \(\theta = 0^\circ\) and \(\theta = 270^\circ\) show a substantial reduction in intensity, likely due to body occlusion and the non-ideal pickup pattern of the microphone in those positions.

Further analysis indicates a gradual decline in signal intensity for frequencies \( f > 20 \) kHz, highlighting the reduced sensitivity of smartwatch microphones at higher ultrasonic frequencies. Additionally, minor variations in spectral energy distribution across angles suggest that reflections from nearby surfaces contribute to secondary wave propagation, further influencing signal reception. These observations emphasize the importance of smartwatch positioning in ultrasonic covert channels, where the function \( S_r(\theta) \) determines the effectiveness of signal reception. Proper alignment enhances data reception, while certain angles—such as \(\theta = 0^\circ\) and \(\theta = 270^\circ\)—naturally degrade transmission efficiency.

The results indicate that smartwatch orientation plays a crucial role in ultrasonic covert communication. To maximize reception efficiency, an attacker can position the smartwatch at angles with a direct line of sight, such as \(\approx 90^\circ\), where \( S_r(\theta) \) reaches its peak. Conversely, defensive strategies can leverage body occlusion effects at larger angles to disrupt covert ultrasonic transmissions.

\subsection{Active Speakers}
Active loudspeakers are self-powered, integrating built-in amplifiers and eliminating the need for external amplification. They are widely used in computers, multimedia systems, and portable audio devices due to their convenience and compact design. In the context of ultrasonic covert communication, the acoustic properties of active speakers must be analyzed to determine their efficiency in transmitting inaudible signals over varying distances.

\begin{table}[h]
	\centering
	\caption{Signal-to-Noise Ratio (SNR) at Various Distances for Different Transmission Rates Using an Active Loudspeaker}
	\label{tab:snr_results}
	\begin{tabular}{|c|c|c|c|c|}
		\hline
		\textbf{Dist. (m)} & \textbf{Raw Sweep (dB)} & \textbf{5 bps (dB)} & \textbf{20 bps (dB)} & \textbf{50 bps (dB)} \\
		\hline
		1  & 35.2 & 32.5 & 30.1 & 27.8 \\
		2  & 32.1 & 29.8 & 27.4 & 24.9 \\
		3  & 29.3 & 27.1 & 24.5 & 22.0 \\
		4  & 26.5 & 24.0 & 22.2 & 19.5 \\
		5  & 23.8 & 21.5 & 19.8 & 17.2 \\
		6  & 21.2 & 19.0 & 17.4 & 15.1 \\
		7  & 18.7 & 16.8 & 15.1 & 13.0 \\
		8  & 16.3 & 14.5 & 12.9 & 11.0 \\
		9  & 14.0 & 12.3 & 11.0 & 9.5 \\
		\hline
	\end{tabular}
\end{table}

Table~\ref{tab:snr_results} presents the measured Signal-to-Noise Ratio (SNR) at distances ranging from \( d = 1 \) m to \( d = 9 \) m for different transmission bit rates. The results show a consistent decline in SNR as transmission distance increases, following the free-space path loss model:

\[
SNR(d) \propto d^{-\gamma}
\]

where \( \gamma \) represents the path loss exponent, influenced by environmental reflections and medium absorption.

The raw ultrasonic sweep maintains the highest SNR across all distances, indicating that unmodulated signals experience less degradation than frequency-shift keying (FSK)-modulated transmissions. Lower bit rates, such as 5 bps, demonstrate better SNR retention, while higher bit rates (50 bps) exhibit greater signal degradation due to increased noise introduced by rapid symbol transitions. At \( d = 9 \) m, the SNR for 50 bps transmission drops to 9.5 dB, suggesting that reliable data transmission becomes challenging at extended distances.

These results indicate that lower transmission rates are more effective for long-distance ultrasonic communication. In contrast, higher bit rates, despite enabling faster data transfer, suffer from rapid SNR degradation, limiting their effectiveness for covert exfiltration in air-gapped environments.

\subsection{Passive Speakers}
Passive loudspeakers rely on external amplifiers to drive their audio output, unlike self-powered active speakers. They are commonly found in home theater systems, high-fidelity audio setups, and professional sound environments, where separate amplifiers provide greater flexibility and control over sound quality. In the context of ultrasonic covert communication, understanding the transmission properties of passive speakers is crucial, as their ability to generate ultrasonic signals depends on both the amplifier characteristics and the speaker’s inherent frequency response.

\begin{table*}[h]
	\centering
	\caption{Signal-to-Noise Ratio (SNR) and Bit Error Rate (BER) for Different Transmission Rates Using Passive Speaker Transmitters}
	\label{tab:snr_passive}
	\begin{tabular}{|c|c|c|c|c|c|c|c|c|}
		\hline
		\multicolumn{3}{|c|}{\textbf{5 bps}} & \multicolumn{3}{c|}{\textbf{20 bps}} & \multicolumn{3}{c|}{\textbf{50 bps}} \\
		\hline
		\textbf{Distance (m)} & \textbf{BER (\%)} & \textbf{SNR (dB)} & \textbf{Distance (m)} & \textbf{BER (\%)} & \textbf{SNR (dB)} & \textbf{Distance (m)} & \textbf{BER (\%)} & \textbf{SNR (dB)} \\
		\hline
		1  & 0\%    & 30  & 1  & 0\%    & 24  & 1  & 8.3\%  & 30  \\
		2  & 0\%    & 10  & 2  & 0\%    & 36  & 2  & 0\%    & 31  \\
		3  & 0\%    & 24  & 3  & 0\%    & 38  & 3  & 0\%    & 29  \\
		4  & 0\%    & 20  & 4  & 0\%    & 30  & 4  & 0\%    & 30  \\
		5  & 0\%    & 20  & 5  & 0\%    & 30  & 5  & 4.16\% & 24  \\
		6  & 0\%    & 18  & 6  & 4.16\% & 15  & 6  & 12.5\% & 26  \\
		7  & 4.16\% & 11  & 7  & 4.16\% & 22  & 7  & 0\%    & 25  \\
		8  & 0\%    & 10  & 8  & 8.33\% & 27  & 8  & 100\%  & 23  \\
		9  & 100\%  & 8   & 9  & 4.16\% & ?   & 9  & 0\%    & 35  \\
		\hline
	\end{tabular}
\end{table*}

Table~\ref{tab:snr_passive} presents the measured Signal-to-Noise Ratio (SNR) and Bit Error Rate (BER) for different transmission bit rates across varying distances using passive speaker transmitters. The results indicate that the SNR decreases as the transmission distance increases, confirming signal attenuation effects. 

The relationship between BER and SNR for binary frequency-shift keying (BFSK) modulation follows the standard Gaussian Q-function:

\[
BER(d) = Q\left(\sqrt{SNR(d)}\right)
\]

where \( Q(x) \) is the complementary cumulative distribution function of the standard normal distribution, defined as:

\[
Q(x) = \frac{1}{\sqrt{2\pi}} \int_x^\infty e^{-\frac{t^2}{2}}\,dt
\]

and \( SNR(d) \) represents the signal-to-noise ratio measured at the receiver located at distance \( d \).

For 5 bps transmission, the BER remains at 0\% across most distances but rises sharply to 100\% at \( d = 9 \) m, indicating complete signal loss. At 20 bps, the BER remains low at shorter distances but becomes more inconsistent beyond \( d = 7 \) m. The 50 bps transmission shows relatively stable SNR performance up to \( d = 6 \) m, but beyond this point, the BER increases significantly, reaching 100\% at \( d = 8 \) m.

\begin{table}[h]
	\centering
	\caption{Sweep Signal SNR Across Different Frequency Ranges and Distances}
	\label{tab:sweep_snr}
	\begin{tabular}{|c|c|c|c|c|c|c|c|c|c|}
		\hline
		\textbf{Freq. (kHz)} & \textbf{1m} & \textbf{2m} & \textbf{3m} & \textbf{4m} & \textbf{5m} & \textbf{6m} & \textbf{7m} & \textbf{8m} & \textbf{9m} \\
		\hline
		18-18.5  & 25 & 35 & 26 & 25 & 23 & 23 & 22 & 25 & 28 \\
		18.5-19  & 26 & 27 & 27 & 25 & 18 & 19 & 19 & 19 & 28 \\
		19-19.5  & 20 & 25 & 19 & 15 & 13 & 11 & 10 & 12 & 14 \\
		\hline
	\end{tabular}
\end{table}

Table~\ref{tab:sweep_snr} confirms that lower frequencies (18-18.5 kHz) exhibit the highest SNR, while frequencies above 20 kHz degrade significantly beyond 5 meters, limiting their viability for long-range ultrasonic covert communication.

\subsection{Laptop Speakers}

Laptops are equipped with built-in speakers that are typically small, low-power, and optimized for mid-range frequencies. Due to space and power constraints, these speakers often have a limited frequency response and lower output levels compared to external speakers. While primarily designed for voice and multimedia playback, they are still capable of generating ultrasonic frequencies, making them relevant for studies on acoustic-based data transmission.

In ultrasonic covert communication scenarios, the characteristics of laptop speakers play a key role in determining the effectiveness of signal propagation. Factors such as output power, directional properties, and frequency response influence the achievable transmission distance and reliability. The following measurements assess the transmission performance of a laptop speaker across different bit rates and distances.

\begin{table*}[h]
	\centering
	\caption{Signal-to-Noise Ratio (SNR) and Bit Error Rate (BER) for Different Transmission Rates Using a Laptop}
	\label{tab:snr_laptop}
	\begin{tabular}{|c|c|c|c|c|c|c|c|c|}
		\hline
		\multicolumn{3}{|c|}{\textbf{5 bps}} & \multicolumn{3}{c|}{\textbf{20 bps}} & \multicolumn{3}{c|}{\textbf{50 bps}} \\
		\hline
		\textbf{Distance (m)} & \textbf{BER (\%)} & \textbf{SNR (dB)} & \textbf{Distance (m)} & \textbf{BER (\%)} & \textbf{SNR (dB)} & \textbf{Distance (m)} & \textbf{BER (\%)} & \textbf{SNR (dB)} \\
		\hline
		1  & 0\%    & 43  & 1  & 0\%    & 40  & 1  & 8.3\%  & 28  \\
		2  & 0\%    & 40  & 2  & 0\%    & 37  & 2  & 0\%    & 40  \\
		3  & 0\%    & 40  & 3  & 0\%    & 35  & 3  & 0\%    & 30  \\
		4  & 0\%    & 34  & 4  & 0\%    & 32  & 4  & 0\%    & 31  \\
		5  & 0\%    & 30  & 5  & 4.1\%  & 33  & 5  & 4.1\%  & 27  \\
		6  & 0\%    & 36  & 6  & 0\%    & 22  & 6  & 12.5\% & 28  \\
		7  & 0\%    & 30  & 7  & 4.1\%  & 28  & 7  & 0\%    & 22  \\
		8  & 0\%    & 28  & 8  & 4.1\%  & 21  & 8  & 100\%  & 30  \\
		9  & 0\%    & 28  & 9  & 0\%    & 32  & 9  & 0\%    & 25  \\
		\hline
	\end{tabular}
\end{table*}

\begin{table}[h]
	\centering
	\renewcommand{\arraystretch}{1.3} 
	\setlength{\tabcolsep}{10pt} 
	\caption{Sweep Signal SNR Across Different Frequency Ranges and Distances}
	\label{tab:sweep_laptop}
	\begin{tabular}{|c|c|c|}
		\hline
		\textbf{Frequency Range (kHz)} & \textbf{SNR at 1m (dB)} & \textbf{SNR at 4m (dB)} \\  
		\hline
		18 - 18.5  & 50 & 41 \\  
		18.5 - 19  & 50 & 40 \\  
		19 - 19.5  & 45 & 35 \\  
		19.5 - 20  & 42 & 36 \\  
		20 - 20.5  & 35 & 24 \\  
		20.5 - 21  & 24 & 29 \\  
		21 - 21.5  & 7  & 5  \\  
		\hline
	\end{tabular}
\end{table}

Table~\ref{tab:snr_laptop} presents the measured Signal-to-Noise Ratio (SNR) and Bit Error Rate (BER) at various transmission rates and distances using a laptop as the transmitter. The results indicate that the SNR remains high at short distances but gradually decreases as the transmission distance increases, following the expected path loss behavior:

\[
SNR(d) \propto d^{-\gamma}
\]

where \( \gamma \) is the path loss exponent, which depends on environmental factors such as reflections and air absorption.

The 5 bps transmission rate maintains an SNR above 28 dB across all distances with 0\% BER, indicating stable and reliable communication over the tested range. At 20 bps, the BER remains low at short distances but starts to degrade at medium distances, reaching 4.1\% at 5, 7, and 8 meters. The 50 bps transmission experiences rapid degradation, with BER increasing to 12.5\% at 6 meters and reaching 100\% at 8 meters, indicating a complete loss of data integrity.

Table~\ref{tab:sweep_laptop} provides SNR measurements for different frequency ranges at distances of 1 m and 4 m. The best performance is observed within the 18-19.5 kHz range, which maintains the highest SNR across both distances. Frequencies above 21 kHz experience significant attenuation, as evidenced by the steep decline in SNR.

These results confirm that lower bit rates, such as 5 bps, offer greater transmission reliability over longer distances, whereas higher bit rates, such as 50 bps, are more susceptible to noise and attenuation. Additionally, the analysis of frequency-dependent SNR suggests that ultrasonic covert channels are more effective when operating within lower frequency bands (18-19.5 kHz), as higher frequencies degrade more rapidly over distance.

\subsection{SNR Comparison Summary}

\begin{figure}[h]
	\centering
	\includegraphics[width=1\linewidth]{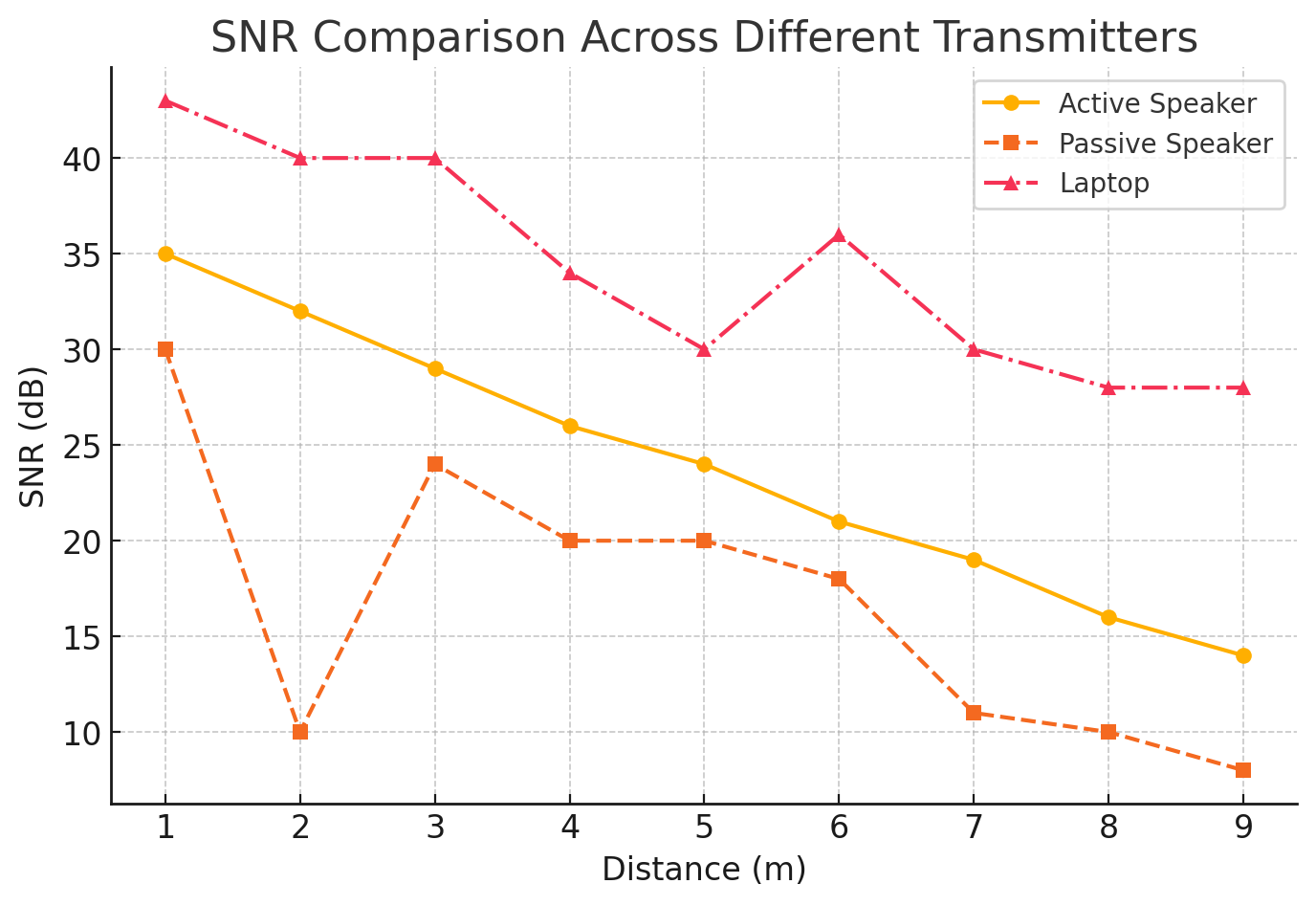}
	\caption{Signal-to-Noise Ratio (SNR) comparison across different transmitters.}
	\label{fig:snr_comparison}
\end{figure}

Figure~\ref{fig:snr_comparison} presents the measured Signal-to-Noise Ratio (SNR) at increasing distances for three different transmitter configurations: active speaker, passive speaker, and laptop. The results indicate that the active speaker maintains the highest SNR across all distances, making it the most effective for long-range ultrasonic communication. In contrast, laptops exhibit stable performance at shorter distances (\( d \leq 4 \) m), but their SNR decreases significantly beyond 6 meters, limiting their suitability for long-range transmission. The passive speaker shows the steepest decline in SNR, with rapid signal degradation, particularly at higher bit rates.

The overall SNR trend follows an inverse power-law relationship:

\[
SNR(d) \propto d^{-\gamma}
\]

where \( \gamma \) is the path loss exponent, influenced by the acoustic propagation characteristics of each transmitter. The data confirms that active speakers provide the most reliable solution for long-distance ultrasonic covert communication, while laptops remain viable for short-range transmission. Passive speakers, however, demonstrate significant attenuation and should not be used beyond 6 meters due to their rapid SNR degradation.

\begin{figure}[h]
	\centering
	\includegraphics[width=1\linewidth]{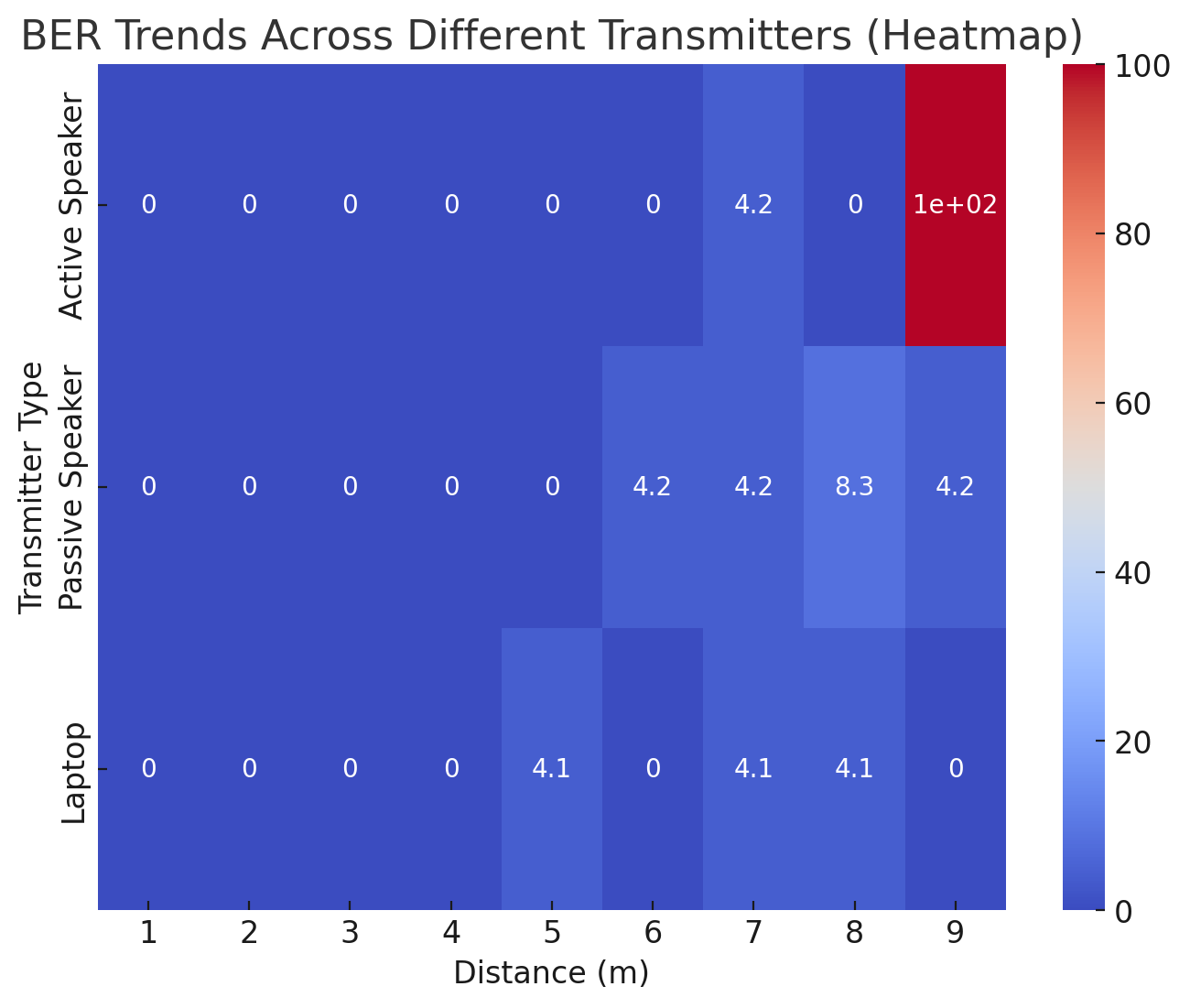}
	\caption{Bit Error Rate (BER) trends across different transmitters.}
	\label{fig:ber_heatmap}
\end{figure}

Figure~\ref{fig:ber_heatmap} presents a heatmap visualization of the bit error rate (BER) as a function of distance (\( d \)) for the three transmitter setups: active speaker, passive speaker, and laptop. The color gradient represents BER trends, where darker regions indicate lower BER (higher transmission reliability), while lighter regions transitioning to red indicate increasing BER and higher data corruption.

The results demonstrate that the active speaker maintains near-zero BER up to 8 meters but experiences a sharp failure at \( d = 9 \) m, where BER reaches 100\%. The passive speaker exhibits gradual BER degradation, with a significant increase beyond \( d = 6 \) m, suggesting that passive sound emission is less effective for long-distance ultrasonic transmission. Laptops display fluctuating BER values between 5 and 8 meters, likely due to environmental factors or hardware variability. However, there is a slight recovery in BER at 9 meters, possibly due to acoustic reflections improving signal reception.

These findings confirm that active speakers are the most reliable for long-range ultrasonic covert communication, while passive speakers and laptops are more suited for distances below 6 meters. Furthermore, higher bit rates are more prone to errors, making 5 bps the most reliable choice for covert ultrasonic exfiltration.

\subsection{Attenuation of Ultrasonic Signals in Smartwatch Receiver}
A thorough analysis of acoustic wave propagation and absorption in different materials is beyond the scope of this paper. However, we provide a fundamental analysis to illustrate the impact of attenuation on ultrasonic covert communication.

When ultrasonic waves propagate through or around physical obstructions, they experience attenuation due to absorption, scattering, and reflection. The extent of attenuation depends on the signal frequency, the properties of the obstructing material, and the effective distance the sound wave travels. In smartwatch-based ultrasonic covert communication, where smartwatches are typically worn on the wrist, the signal path between the transmitter (e.g., a workstation speaker) and receiver (smartwatch microphone) is often influenced by these attenuation effects.

The attenuation of an ultrasonic signal passing through an obstruction can be modeled as:

\begin{equation}
	A(f, d) = A_0 e^{-\alpha(f) d}
	\label{eq:attenuation}
\end{equation}

where:
\begin{itemize}
	\item \( A(f, d) \) is the attenuated sound intensity at frequency \( f \) after passing through the material.
	\item \( A_0 \) is the initial sound intensity before attenuation.
	\item \( \alpha(f) \) is the frequency-dependent attenuation coefficient (in \(\text{dB/cm}\)), which varies based on the type and density of the obstructing material.
	\item \( d \) is the effective thickness of the obstruction the sound wave travels through (in cm).
\end{itemize}

For ultrasonic frequencies (18 kHz and above), soft materials such as fabrics and human tissue typically exhibit attenuation coefficients between \( 0.1 \) to \( 0.5 \) \(\text{dB/cm}\), while denser materials like glass, plastic, and metal can introduce significantly higher attenuation. As a result, ultrasonic waves used in smartwatch-based covert channels may suffer degradation if obstructions are present in the transmission path.

The level of attenuation varies based on the propagation path:

\begin{itemize}
	\item \textbf{Direct Line-of-Sight:} When the smartwatch microphone has an unobstructed path to the transmitting speaker, attenuation is minimal, typically between 2 and 5 dB at a distance of 1 meter, ensuring high reception quality.
	
	\item \textbf{Partial Obstruction:} When an object partially blocks the signal, moderate attenuation occurs, reducing signal strength while still allowing data reception. A partially occluded path may introduce an additional 10 to 15 dB loss.
	
	\item \textbf{Complete Obstruction:} When a dense material fully blocks the direct path, attenuation increases significantly, potentially exceeding 25 to 30 dB, making reception unreliable unless reflections or secondary propagation paths compensate for the loss.
\end{itemize}

The attenuation effect can also be quantified using the logarithmic path loss model:

\begin{equation}
	A_{\text{dB}}(f, d) = A_0 - \alpha(f) d
\end{equation}

where \( A_{\text{dB}}(f, d) \) denotes the received signal strength at frequency \( f \) over a transmission distance \( d \), \( A_0 \) represents the initial transmission power, and \( \alpha(f) \) is the frequency-dependent attenuation coefficient (in dB per meter).

For example, assuming a 19 kHz signal transmitted over a distance of 1.5 meters with an initial power of 60 dB and an attenuation coefficient of 5 dB per meter. 
When the path is partially obstructed, such as by a fabric sleeve or a user’s wrist, an additional 10 dB attenuation may reduce the received signal to 42.5 dB. If the smartwatch is fully obstructed, such as when worn on the opposite wrist with the torso blocking the signal, attenuation may exceed 30 dB, reducing the received strength to 22.5 dB, making signal recovery challenging.
%

\subsection{Interference from Keyboard Typing}
As smartwatches are frequently worn while using computers, the close proximity of the smartwatch microphone to the keyboard raises concerns about potential interference between typing noise and ultrasonic transmissions. However, spectral analysis indicates that this interference is minimal due to the distinct separation of frequency bands.

\begin{figure}[h]
	\centering
	\includegraphics[width=1\linewidth]{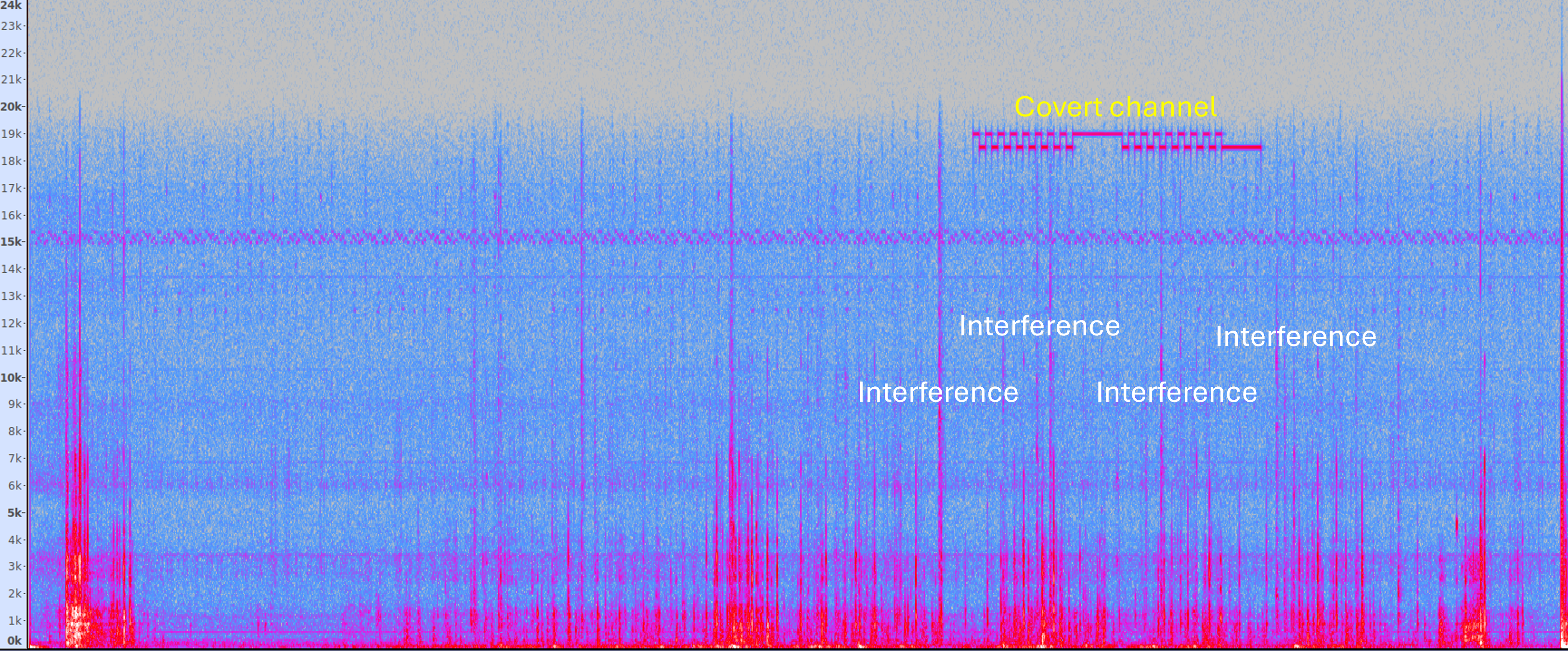}
	\caption{Spectrogram analysis of ultrasonic transmission (18.5–19 kHz) with simultaneous keyboard typing. Keyboard noise spans a broad frequency range but shows minimal interference with the ultrasonic signal.}
	\label{fig:spectrogram}
\end{figure}

To assess the impact of keyboard noise, an ultrasonic transmission within the 18.5–19 kHz range was recorded while a user was actively typing on a standard mechanical keyboard. The spectrogram analysis of the recorded signal, presented in Figure~\ref{fig:spectrogram}, illustrates that acoustic noise generated by keystrokes spans a broad frequency range across the spectrum but is most prominent in the lower and mid-frequency regions. In contrast, the ultrasonic transmission remains confined to frequencies above 18 kHz. This spectral separation ensures that the smartwatch receiver can demodulate and decode the ultrasonic signal without significant disruption from keyboard activity.

\section{Mitigation}
Mitigating the risks associated with ultrasonic covert channels in smartwatches and other recording-capable devices necessitates a multi-faceted security approach. Effective countermeasures must balance security enforcement with operational feasibility to minimize the likelihood of unauthorized data exfiltration while maintaining usability.

Restricting or prohibiting the use of smartwatches and similar audio-capable wearables in sensitive environments is a direct mitigation strategy. Organizations have implemented such policies to enhance security by reducing the potential for covert communication. However, enforcement presents significant challenges, as users may attempt to circumvent restrictions, and continuous monitoring introduces additional resource overhead \cite{govtech2023}. In professional settings, a blanket ban on such devices may also hinder productivity, necessitating a more nuanced approach.

Deploying ultrasonic monitoring systems provides a real-time detection mechanism to identify unauthorized ultrasonic transmissions. However, these systems may yield false positives due to environmental noise, as common devices such as motion sensors, industrial machinery, and even certain consumer electronics can emit ultrasonic frequencies that may be misclassified as covert signals \cite{carrara2015acoustic}. These systems utilize specialized sensors to scan for anomalous ultrasonic frequencies indicative of covert communication attempts. However, their deployment involves high implementation costs and the potential for false positives due to the prevalence of ultrasonic signals in everyday devices \cite{cyansec2023}. Furthermore, sophisticated adversaries may leverage evasion tactics, such as frequency shifting and signal obfuscation, to bypass detection.

Ultrasonic jamming represents another potential countermeasure, wherein intentional ultrasonic noise is emitted to disrupt unauthorized transmissions. While this approach effectively prevents covert data exfiltration, it poses risks of unintended interference with legitimate ultrasonic-dependent systems, including medical and industrial sensors. Additionally, the long-term effects of continuous ultrasonic jamming remain uncertain, raising concerns about regulatory compliance \cite{mavroudis2017privacy}. Systems such as SoniControl provide users with the ability to detect and block ultrasonic tracking, contributing to enhanced security in mobile environments \cite{mavroudis2017sonicontrol}.

An advanced security mechanism involves integrating ultrasonic firewalls within computers. These firewalls operate at the software level (e.g., the OS kernel audio driver), dynamically analyzing incoming and outgoing acoustic signals to identify and filter suspicious ultrasonic frequencies. By leveraging signal processing algorithms, ultrasonic firewalls can effectively differentiate between benign and malicious transmissions \cite{mavroudis2017sonicontrol}. Combining such defenses with existing endpoint security solutions can provide comprehensive protection against ultrasonic-based attacks.

A more restrictive yet highly effective mitigation strategy is \textit{audio-gapping}, wherein audio hardware components, such as microphones and speakers, are physically removed or disabled in air-gapped and highly secure environments. This method eliminates the potential for acoustic-based covert channels by preventing both transmission and reception of ultrasonic signals \cite{airgap2025}. 

\section{Conclusion}
In this paper, we explore a smartwatch-based ultrasonic cyberattack capable of covert data exfiltration from air-gapped environments. We present and evaluate \textit{SmartAttack}, an attack method that exploits the built-in microphones of smartwatches to receive and decode ultrasonic signals in the 18–22 kHz frequency range. Through extensive experimentation, we demonstrate that this attack can successfully transmit data over distances exceeding 6 meters, achieving data rates of up to 50 bits per second.

Our analysis highlights smartwatch-specific factors that affect signal reception, including wrist movement, signal attenuation due to the human body, and the directional limitations of built-in microphones. These factors introduce both operational constraints and advantages compared to previously studied ultrasonic receivers, such as smartphones.

\balance 
\bibliographystyle{IEEEtran}
\bibliography{SMARTWATCHES}

\end{document}